\documentclass[aps,prl,twocolumn,longbibliography]{revtex4-2}

\usepackage{amssymb}
\usepackage{amsbsy}
\usepackage{amsmath}
\usepackage{graphicx}
\usepackage{graphics}
\usepackage{setspace}
\usepackage{array}
\usepackage{color}
\usepackage{fontenc}
\usepackage{textcomp}
\usepackage{bm}
\usepackage{float}
\usepackage[bookmarks=false,linkcolor=blue,urlcolor=blue,colorlinks,citecolor=blue]{hyperref}
\usepackage[normalem]{ulem}

\newcommand{\vex}[1]{\bm{\mathrm{#1}}}

\newcommand{\blue}[1]{{\color{blue}{#1}}}

\newcommand{\bsub}{\begin{subequations}}
\newcommand{\esub}{\end{subequations}}

\graphicspath{{../Figures/}}

\begin{document}
\title{Altermagnetic Routes to Majorana Modes in Zero Net Magnetization}
\author{Sayed Ali Akbar Ghorashi$^{1}$}\email[Correspondence\,to:\,]{sayedaliakbar.ghorashi@stonybrook.edu}
\author{Taylor L. Hughes$^2$}
\author{Jennifer Cano$^{1,3}$}
\affiliation{$^1$Department of Physics and Astronomy, Stony Brook University, Stony Brook, New York 11794, USA}
\affiliation{$^2$Department of Physics and Institute for Condensed Matter Theory,  University of Illinois at Urbana-Champaign, IL 61801, USA}
\affiliation{$^3$Center for Computational Quantum Physics, Flatiron Institute, New York, New York 10010, USA}

\date{\today}

\newcommand{\be}{\begin{equation}}
\newcommand{\ee}{\end{equation}}
\newcommand{\bea}{\begin{eqnarray}}
\newcommand{\eea}{\end{eqnarray}}
\newcommand{\h}{\hspace{0.30 cm}}
\newcommand{\vs}{\vspace{0.30 cm}}
\newcommand{\n}{\nonumber}

\begin{abstract}
We propose heterostructures that realize first and second order topological superconductivity with vanishing net magnetization by utilizing altermagnetism.
Such platforms may offer a significant improvement over conventional platforms with uniform magnetization since the latter suppresses the superconducting gap.
We first introduce a 1D semiconductor-superconductor structure in proximity to an altermagnet which realizes end Majorana zero modes (MZMs) with vanishing net magnetization.
Additionally, a coexisting Zeeman term 
provides a tuning knob to distinguish topological and trivial zero modes.
We then propose 2D altermagnetic platforms that can realize chiral Majorana fermions or higher order corner MZMs. Our work paves the way towards realizing Majorana boundary states with an alternative source of time-reversal breaking and zero net magnetization.
\end{abstract}
\maketitle

Realizing a topological superconductor (TSC) has been a driving force in the development of topological phases of matter, motivated by the goal of producing Majorana states for quantum computation \cite{sato2017topological,Chiu2016,reviewQC}.
While intrinsic realizations remain elusive,
much research has been devoted to engineering hybrid platforms that realize TSCs in 1D nanowires and 2D heterostructures \cite{Majoran1,PhysRevB.88.020407,nadj2014observation, QHZchiralTSC, PhysRevB.92.064520}. \\
\indent The three main ingredients in these platforms are
superconductivity, spin-orbit coupling, and a time-reversal breaking element  (e.g., an applied magnetic field, an adjacent ferromagnet, or magnetic adatoms).
However, despite intense efforts, a definitive realization of Majorana qubits is still lacking \cite{das2023search,DisorderfieldMZM1,kayyalha2020absence,aghaee2022inas}.
The main challenges are: (1) the detrimental effect of disorder, which can create accidental states below the superconducting gap that are difficult to distinguish from protected Majorana zero modes (MZMs),  and (2) control of the proximity-induced superconducting gap, which is typically suppressed by the time-reversal breaking element.
\begin{figure}[tb!]
    \centering
\includegraphics[width=0.45\textwidth]{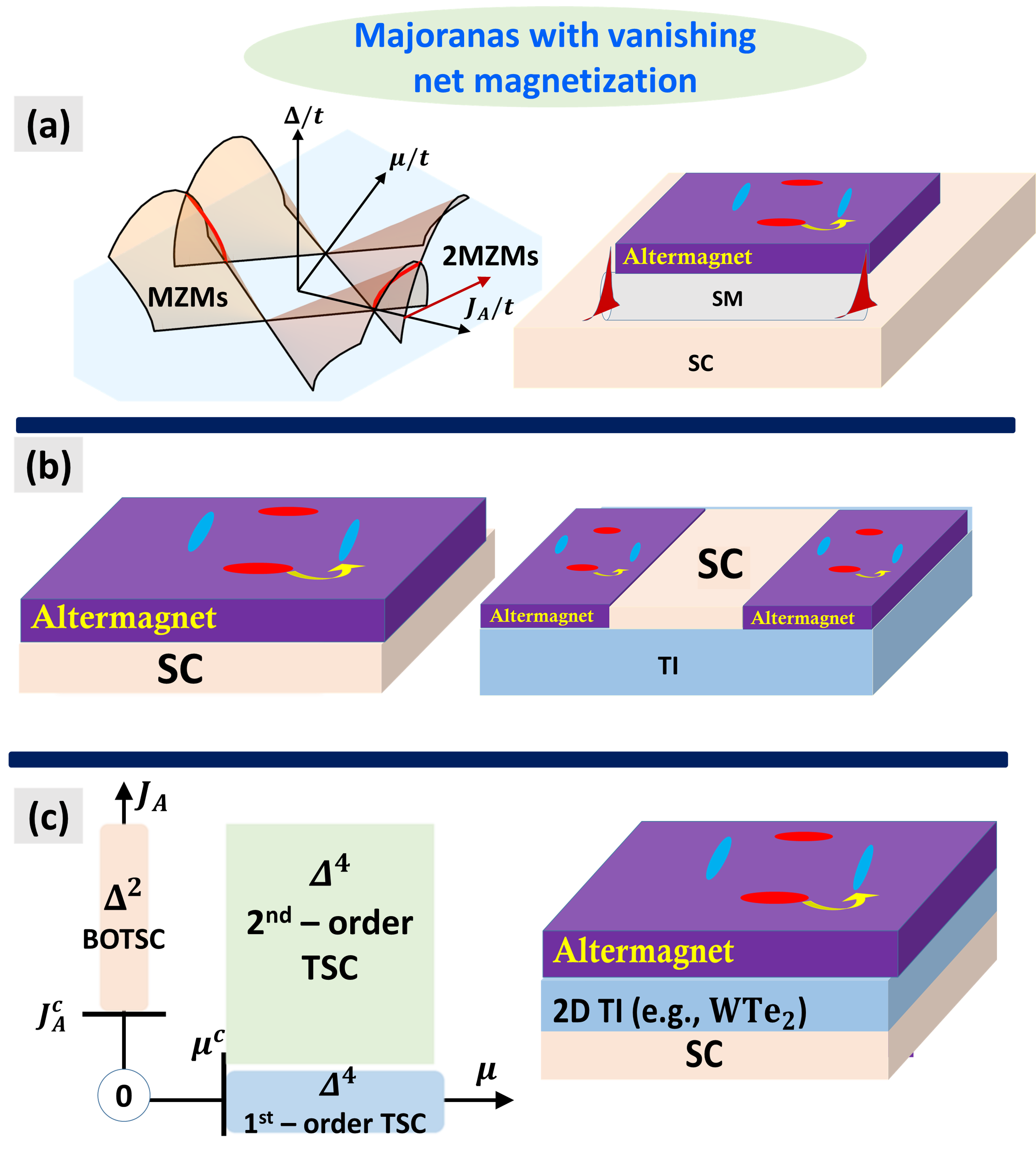}
    \caption{Proposed setups using altermagnets to realize topological superconductivity with vanishing net magnetization. (a)   A 1D altermagnet/ semiconductor(SM)/ superconductor(SC) wire hosts single (shaded area) and multiple (along $\mu/t=0$; red line shows the boundary) MZMs. (b) 2D altermagnet/SC and altermagnet/SC/3D topological insulator (TI) heterostructures generate chiral Majorana modes. (c) An altermagnet/SC/2D TI heterostructure hosts corner MZMs.}
    \label{fig:adpic}
\end{figure}

In this work, we introduce a new platform that can address both problems by realizing MZMs in a system with \emph{vanishing net magnetization}, which allows further tunability with a Zeeman field that can distinguish disorder-induced subgap modes from topological MZMs.
The new ingredient is altermagnetism.
Altermagnets are a class of collinear antiferromagnets with a momentum dependent magnetic order parameter \cite{altermagnet1,altermagnet2} that have recently attracted attention\cite{PhysRevLett.130.216701,PhysRevB.107.L100418,ouassou2023josephson, zhang2023finite,papaj2023andreev,zhou2023crystal}.
Crucially, despite their vanishing magnetization, altermagnets can generate sizeable spin-splitting, which changes sign in different regions of the Brillouin zone (BZ).
For the sake of concreteness, we here focus on a $d$-wave magnetic order parameter, which describes the order of, e.g., RuO$_2$ \cite{altermagnet2}.
However, the route we propose is quite general and we expect similar phenomena for order parameters with other angular momenta.

In the following, we show that 1D and 2D TSCs hosting MZMs and chiral Majorana fermions (CMFs), respectively, can be realized in heterostructures where altermagnetic order replaces the time-reversal breaking element.
In addition to MZMs and CMFs, this platform can also realize higher order TI (HOTI) and higher order TSC phases in 2D.
Importantly, we also show that in 1D a weak uniform magnetization/Zeeman field provides a tuning knob that can help distinguish disorder-induced mid-gap states from protected MZMs.

\emph{\blue{Majorana nanowire: Majorana zero modes without magnetic field}}.---
We first consider a 1D semiconducting (SM) nanowire on the surface of an altermagnet and in proximity to an $s$-wave superconductor (SC), as shown in Fig.~\ref{fig:adpic}(a). The  Bogoliubov-de Gennes (BdG) Hamiltonian of the nanowire can be written as 
\begin{align}\label{wire}
    h(k)=\left[\epsilon(k)+\lambda_R \sin(k) \sigma^2 + J_{A}\cos(k)\sigma^3\right]\tau^3+\Delta\tau^2\sigma^2,
\end{align}
where $\epsilon(k)=t\cos{(k)}-\mu$, $\tau$ and $\sigma$ are Pauli matrices in Nambu and spin space respectively, $t$ is the hopping strength,  $\lambda_R$ is the strength of the Rashba spin-orbit coupling, and $J_A$ and $\Delta$ denote the proximity-induced altermagnetism and superconducting pairing, respectively.
We assume the altermagnetism in the wire is induced from proximity to an altermagnet that has a $d$-wave order order parameter $\left[ \cos(k_x)-\cos(k_y) \right]\sigma^3.$
We expect the strength of the proximity induced altermagnetism $J_A$ to depend on the relative orientation of the wire and the altermagnet, but the form of the coupling should be independent of orientation as long as it is not fine-tuned to lie along the nodal lines of the altermagnetic order.

\begin{figure}[tb!]
    \centering
    \includegraphics[width=0.49\textwidth]{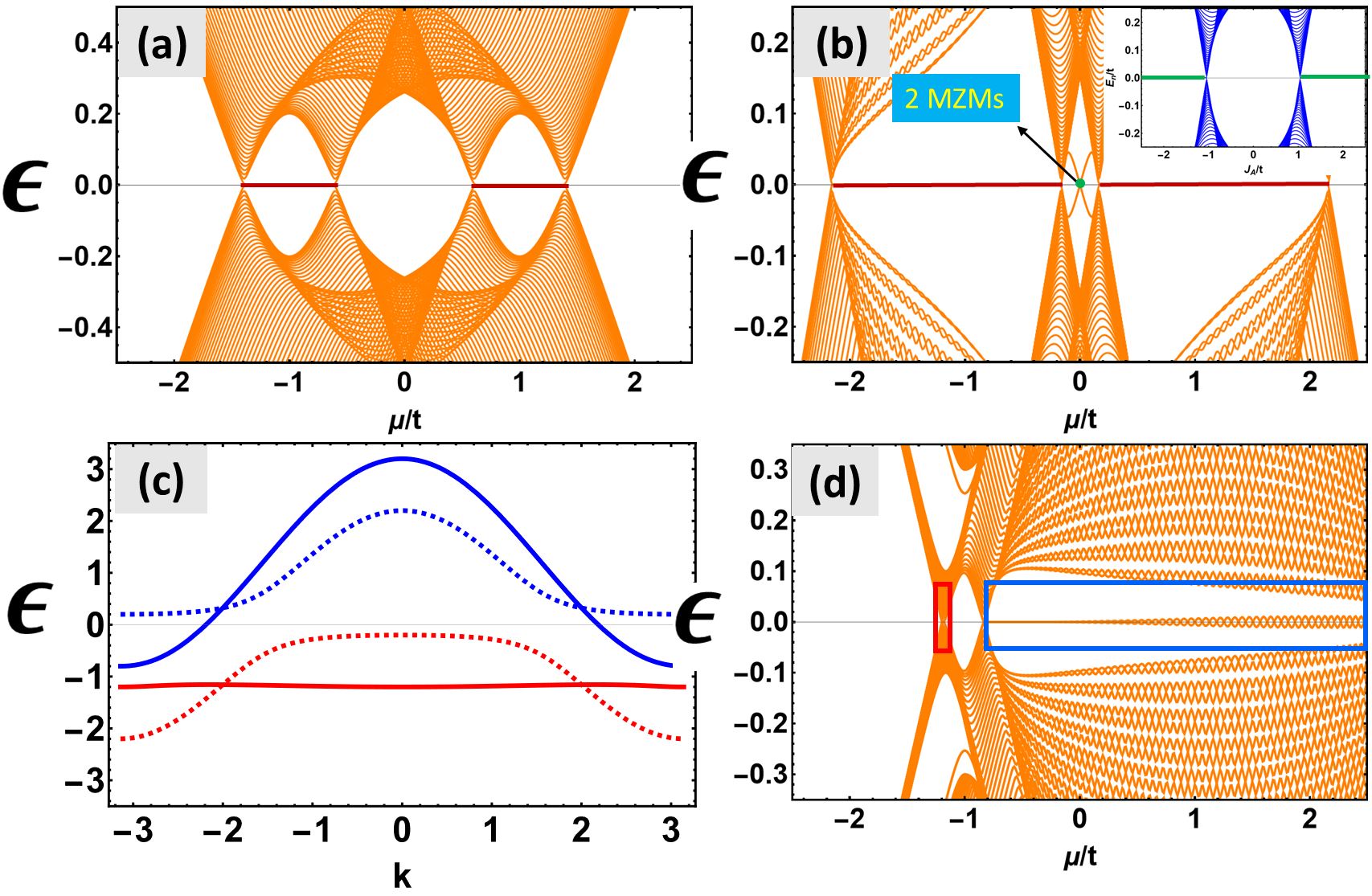}
    \caption{Spectra of a nanowire heterostructure with vanishing net magnetization. (a) Finite-size spectrum of SC-proximitized nanowire vs $\mu$ with $\lambda_R=0.5 t,\,J_A=0.5 t,\, \Delta=0.3 t$; red lines show MZMs. (b) Finite-size spectrum with $\lambda_R=0.5 t,\,J_A=1.2 t,\,\Delta=0.3 t$; inset spectrum vs $J_A$ ($\mu=0$) where green lines show two MZMs.
    (c) Bulk spectrum of nanowire with $\lambda_R=0.6 t,\,J_A=1.2 t,\,J_z=t$ (solid) and $J_z=0$ (dashed) and (d) the corresponding finite-size spectra for solid lines in (c) vs $\mu$ with $\Delta=0.1 t$ showing MZMs only when the dispersive blue band in (c) is doped (blue box in (d)).}
    \label{fig:nanowire}
\end{figure}

Generically the Hamiltonian is in class D, i.e., it has a $\mathbb{Z}_2$
topological classification \cite{schnyder2008classification,schnyder2009classification,kitaev2009periodic} and is topologically non-trivial when $\sqrt{\Delta^2+(t-\mu)^2} <J_A <\sqrt{\Delta^2+(t+\mu)^2}$. In this regime, despite vanishing net magnetization,
this Hamiltonian hosts MZMs at the ends of a finite-sized nanowire, as shown in Fig.~\ref{fig:nanowire}(a).
These MZMs are similar to those that arise in uniform magnetization: they arise when the nanowire is metallic with an odd number of partially filled subbands that are gapped by the superconductivity. By eliminating the net magnetization, the altermagnetic heterostructure we propose may offer a significant improvement over MZM platforms with uniform magnetization since the latter suppresses the superconducting gap \cite{PhysRevB.96.075161, DisorderfieldMZM1}.
Indeed, despite vanishing net magnetization, the altermagnet term is effective in replacing the ferromagnet because it is non-vanishing on generic Fermi surfaces where $\cos k\neq 0$; this will turn out to be the crux of designing TSCs with altermagnets, as will become more clear below when we consider models in two dimensions.

While our altermagnet can directly reproduce the MZMs seen in ferromagnetic systems, we also find phenomena that distinguish the two time-reversal breaking elements. For example, in Fig.~\ref{fig:nanowire}(b) we show the open boundary spectrum as a function of $\mu$ for $|J_A|>|t|, \alpha=0.5t, \Delta=0.3t.$ In this regime the altermagnetism, combined with the Rashba term, generates a non-trivial Su-Schrieefer-Heeger\cite{su1979solitons} phase in the normal state, i.e., the two normal-state bands each have a $\pi$ Berry phase.
For $\mu\sim 0,$ i.e., when $\mu$ is in or near the insulating gap, the resulting superconducting phase will have a \emph{pair} of MZM end states arising from the single complex fermion bound state in the normal state. Generically, these MZMs couple and open a gap.
However, if $\mu$ is tuned so that the edge modes in the normal state are at zero energy, which occurs at $\mu = 0$ for the Hamiltonian in Eq.~(\ref{wire}), then the resulting MZMs will be degenerate, as seen in Fig. \ref{fig:nanowire}(b). If one tunes $|t|$ larger then $|J_A|$, there will generically be an indirect gap closing for $\mu\sim 0$ which does not generate the pair of boundary MZMs. We note this physics does not occur in uniform magnetization, for which the normal phase would be a trivial 1D insulator when magnetism is stronger than $t$.

Importantly, our model in Eq. \eqref{wire} has a useful sensitivity to an applied magnetic Zeeman field. Explicitly, consider the effect of a weak external magnetic field, which induces a Zeeman term $J_Z \hat{\mathbf{B} } \cdot \bm{\sigma}$ in the normal state. The application of a Zeeman field $B_z,$ which is parallel to the altermagnet axis, can tune the bandwidth of the subbands, e.g., as shown in Fig. \ref{fig:nanowire}(c). We find a generic regime where a Zeeman field parallel to the altermagnet axis, in the presence of Rashba spin orbit coupling and altermagnetism, can suppress the bandwidth and flatten one of the spin-split subbands. As a consequence, the open-boundary spectra in Fig. \ref{fig:nanowire}(d) exhibit a dramatic asymmetry as a function of $\mu$ where there are narrow and wide topological regions corresponding to gating into the narrow or wide subbands in Fig. \ref{fig:nanowire}(c). When $\mu$ is in the narrow band, the presence of MZMs can be easily tuned using a weak Zeeman field. Since spurious subbgap states created by non-magnetic disorder will not be as sensitive to the applied field, one can use this effect to distinguish non-topological and topological bound states, e.g., the topological ones can be easily created or destroyed by tuning the Zeeman field. Furthermore, since a subband can be essentially flattened, the relative effects of interactions can be tuned and may lead to new phenomena.

We note that Zeeman fields perpendicular to the altermagnetic axis do not show a similar effect, i.e., they do not appreciably tune the band width. We can compare this to the conventional case where a Zeeman field in the (possibly two directions) perpendicular to the Rashba term will modify the spin-split subband gaps, but will not exhibit such strong bandwidth tuning.
\blue{\emph{Chiral Majorana fermions with zero net magnetization}}.---We now study the two-dimensional altermagnetic heterostructures shown in Fig.~\ref{fig:adpic}(b), which can host chiral modes and, in the presence of superconductivity, CMFs.\\
\textbf{Approach I: Altermagnet/3D TI interfaces}
There are many examples of engineered heterostructures involving 3D time-reversal-invariant topological insulators (3D TIs) in proximity to superconductors and time-reversal breaking magnetic elements \cite{QHZchiralTSC,PhysRevLett.100.096407,PhysRevB.92.064520}. In previous work the magnetic element has been a ferromagnet, but we can straightforwardly apply these results to altermagnets. The key requirement, as we mentioned earlier, is that the topological surface states must sit at momenta where the altermagnetic order is non-vanishing and can open a gap. To be explicit, consider depositing an altermagnet with order parameter $(\cos k_x -\cos k_y)\sigma^3$ on the $\hat{z}$-normal surface of a 3DTI. In the case of perfect alignment, where the $k_x,k_y$ axes of the altermagnet and 3DTI surface align, the altermagnetism can open gaps of opposite sign on surface Dirac cones at the $X=(\pi,0)$ and $ Y=(0,\pi)$ points of the surface Brillouin zone, and leave cones at the $\Gamma =(0,0)$ and $ M=(\pi,\pi)$ points gapless. (If the altermagnetism persisted in the bulk it could yield a 3D HOTI with chiral hinge modes \cite{PhysRevB.96.245115,schindler2018higher}.)

Now consider the case of $N_{X\pm} (N_{Y\pm})$ Dirac cones at the $X$- ($Y$-)point with helicity $\pm$ and no Dirac cones at $\Gamma$ or $M$.
At a domain wall/interface where the altermagnetism flips sign, the Chern number will change by $|\Delta C|=|N_{X+}-N_{X-}-N_{Y+}+N_{Y-}|$,
and hence there will be a net $|\Delta C|$ chiral modes at the interface.
One might create a domain wall by flipping the altermagnet axis via an applied field.
An alternate approach is to use the anisotropy of the altermagnet to create a domain wall: indeed, rotating the altermagnet by $\theta$ such that $\pi/4<\theta<3\pi/4$ switches the sign of the order parameter at the $X$ and $Y$ points.
Hence, chiral modes can be found on engineered or naturally occurring domains of the altermagnetism.
Using the same logic, a 3DTI with altermagnet layers on top and bottom (each with the same sign) will be a Chern insulator whose Chern number has magnitude $|\Delta C|$.

If the altermagnet is misaligned by an angle $\theta\neq n\pi/2$ for integer $n,$ then several things may occur: (i) it may also open a gap for cones at the $M$-point, and (ii) there will be incommensurate effects since the Brillouin zone rotated by $\theta$ will not match that of the 3D TI. We leave further discussion of generic alignments to future work.

We can also imagine an interface between an altermagnet and an $s$-wave SC on the surface of a 3DTI. Under the same surface state assumptions above, there is a net number of $|\Delta C|$ CMFs at the interface.  Similarly, a 3DTI with an altermagnet on one surface and an $s$-wave SC on the other yieds a chiral SC with a superconducting Chern number of magnitude $|\Delta C|.$ As a related effect, if one made an interface between two 3DTI phases (A) and (B) with different $N_{X/Y,\pm}$ then even a single-domain altermagnet that covered both 3DTIs could generate a net number of chiral interface states $\tfrac{1}{2}|\Delta C_B-\Delta C_A|$. We conclude that the ferromagnetic element can be replaced by an altermagnet as long as the surface states reside at momenta off the nodal lines/surfaces of the altermagnet.

\textbf{Approach II: bulk altermagnet/superconductor heterostructures} We simplify the platform by removing the 3DTI and considering altermagnet/SC heterojunctions as in Fig.~\ref{fig:adpic}(b). We use a model consisting of an altermagnet layer coupled to a SC layer with $s$- and/or $d$-wave singlet pairing. The Hamiltonian is given by:
\begin{align}\label{HI}
    H_{I}(\mathbf{k}) &=[\epsilon(\vex{k})-\mu]\tau^3+\lambda_{R}(\vex{k})+J_A(\vex{k})+\Delta(\vex{k})\tau^2\sigma^2,
\end{align}
where $\epsilon(\vex{k})=t(\cos(k_x)+ b_1 \cos(k_y))$, $\lambda_{R}(\mathbf{k})=\lambda_{R}(\sin(k_x)\tau^3\sigma^2-\sin(k_y)\tau^0\sigma^1)$, $J_A(\mathbf{k})=J_A(\cos(k_x)-b_2 \cos(k_y))\tau^3\sigma^3$, $\Delta(\vex{k})=\Delta_0+\Delta_1(\cos(k_x)+b_3 \cos(k_y))$; $t,\,\lambda_{R},\,J_{A}, \Delta_{0,1}$ represent hopping, Rashba spin-orbit coupling, altermagnet strength, and superconducting pairing amplitudes, and we have included the parameters $0<b_{1,2,3}<1$ to represent anisotropic distortions along the $y$-direction for hopping, magnetic, and superconducting strength respectively. The parameters $b_{1,2}$ are generically independent; for example, hopping anisotropy where $b_1\neq 1$ does not always change the magnetic structure \cite{vsmejkal2020crystal}. Since we find $b_2\neq 1$ does not qualitatively change the results, we set $b_2=1, t=1$ in the following. Interestingly, in the absence of superconductivity, an altermagnet can be gapped out and turn into a Chern insulator by applying a weak uniform magnetization. However, because in this work our focus is on TSCs with zero net magnetization we do not consider any uniform external magnetization. We now discuss three important limits (I-III) of Eq.~\eqref{HI}.
\begin{figure}[tb!]
    \centering
    \includegraphics[width=0.41\textwidth]{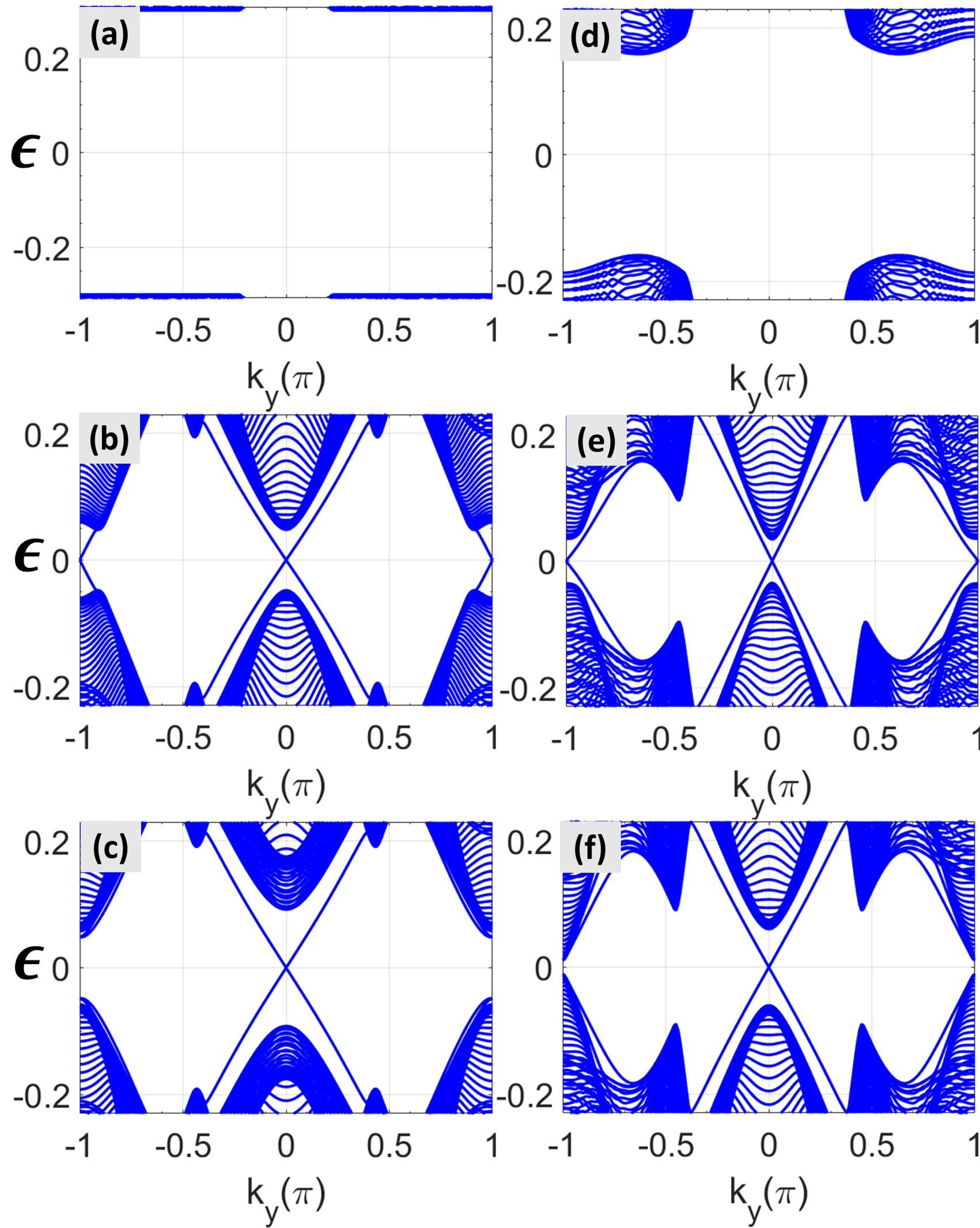}
    \caption{The $x$-normal edge spectrum of the altermagnet/SC heterojunction in Eq.~\eqref{HI} with $s$ (a-c) and $s_{\pm}-$wave (d-f) pairings. (a) $\Delta_0=0.3,\, b_1=1,\,J_A=0$, (b) $\Delta_0=0.3,\, b_1=1,\,J_A=0.3$, (c) $\Delta_0=0.3,\, b_1=0.85,\,J_A=0.3$, (d) $\Delta_0=0.4,\,\Delta_1=0.3,\, b_1=1,\,b_{3}=1,\,J_A=0$, (e) $\Delta_0=0.4,\,\Delta_1=0.3,\,b_1=1,\,b_{3}=1,\,J_A=0.3$, (f) $\Delta_0=0.4,\,\Delta_1=0.3,\, b_1=1,\,b_{3}=0.8,\,J_A=0.3$. For all plots, $\mu=-0.4, \lambda_R=0.3$.}
    \label{fig:swave}
\end{figure}

(I) \underline{\emph{Altermagnet/$s$-wave SC}}.
First consider the case of isotropic $s$-wave pairing $\Delta_0\neq 0$, $\Delta_1 = 0$. Figs.~\ref{fig:swave}(a,b) compare the open boundary spectra of Eq.~\eqref{HI} for $J_A=0$ and $J_A\neq 0$, with $b_1=1, \lambda_R=0.3, \Delta_0=0.3, \mu=-0.4$.  For $J_A=0$ the normal state has two, Rashba-like Fermi surfaces around the $\Gamma$-point, yielding a trivial superconducting phase, i.e., the bulk is fully gapped and no surface states. In contrast, for $J_A=0.3$, the normal state has one Fermi surface around $\Gamma$ and another around $M$, yielding a weak TSC with vanishing Chern number, but hosting two Majorana edge modes of opposite chirality protected by translation symmetry, e.g., on an $x$-normal edge there is one mode each at $k_y=0,\pi$. Adding a slight crystal anisotropy, e.g., $b_1=0.85$ a strong chiral TSC with SC Chern number $|\mathcal{N}|=1$ appears despite the vanishing net magnetization (see Fig.~\ref{fig:swave}(c)). The anisotropy of the altermagnet forces the CMFs to reside at different momenta for edges related by a $\pi/2$ rotation, e.g., we find a chiral mode at $k_y=0$ on an $x$-normal edge, but at $k_x=\pi$ for a $y$-normal edge. This configuration is reversed by changing the sign of $\mu$ (relative to half-filling); hence a momentum-resolved spectroscopy measurement would observe alternating profiles between $x$- and $y$-normal edges as a function of doping or gating.

(II) \underline{\emph{Altermagnet/ anisotropic $s$+$s_{\pm}$-wave SC}}. Similar topological phases for an isotropic crystal are obtained by resorting to a slightly anisotropic $s_{\pm}$ pairing, i.e., $\Delta_0, \Delta_1\neq 0,\,0<b_3< 1$. Figs.~\ref{fig:swave}(d,e) compare the open boundary spectra ($x$-direction open) without and with $J_A$ respectively. Similar to limit (I), a weak TSC is generated with isotropic $s_{\pm}$ and non-vanishing altermagnetism (see Fig. \ref{fig:swave}(e)) while the case with vanishing altermagnetism is trivial. Including a small anisotropy in the pairing, $b_3\neq 1$, yields a chiral TSC with $|\mathcal{N}|=1$, Fig.~\ref{fig:swave}(f). In contrast to (I), the momenta of the Majorana modes on each edge are correlated to the sign (phase) difference between $\Delta_0$ and $\Delta_1$ instead of the sign of $\mu$.

\begin{figure}[tb]
    \centering
    \includegraphics[width=0.41\textwidth]{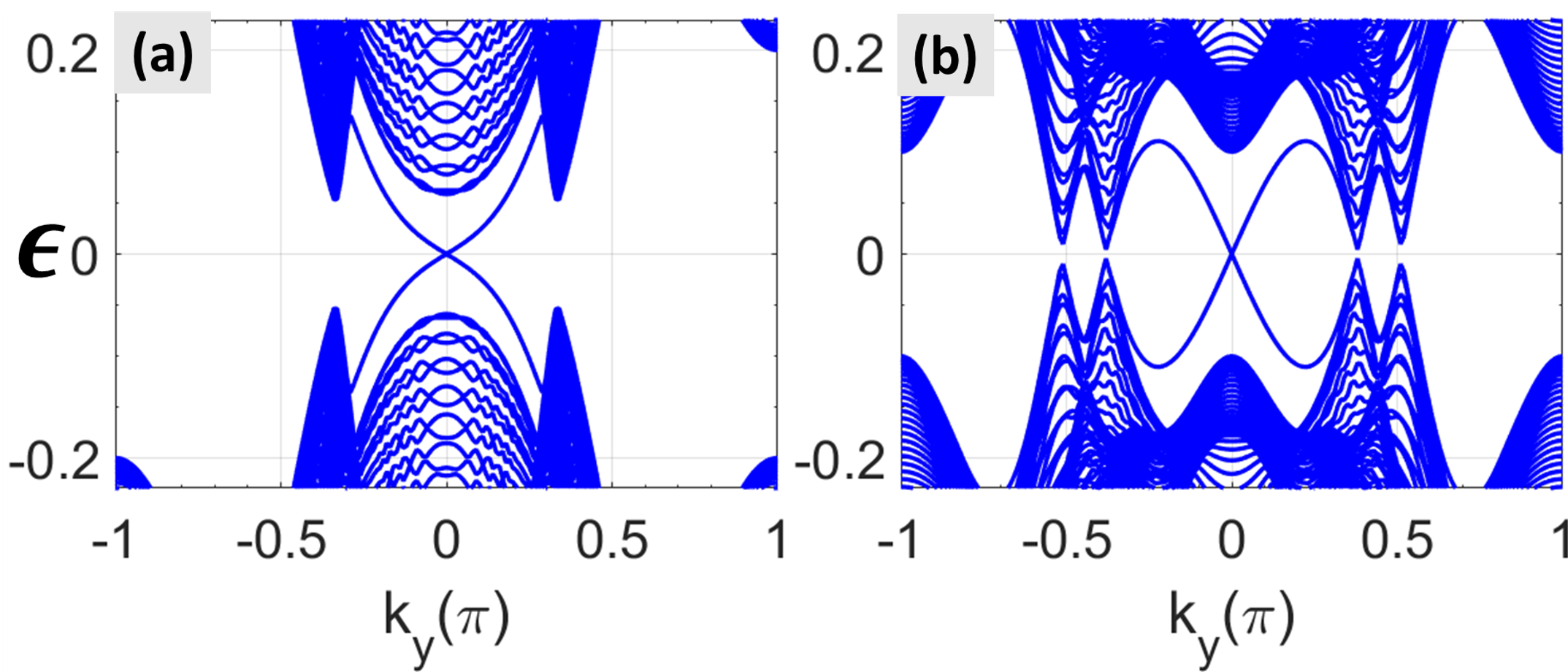}
    \caption{The $x$-normal edge spectrum of the altermagnet/SC heterojunction in Eq.~\eqref{HI} with $s+d$-wave pairing. (a) $\Delta_0=0.4,\,\Delta_1=0.2,\,\mu=0,\,b_1=1,\,J_A=0.3$, (b) $\Delta_0=0.1,\,\Delta_1=0.3,\,\mu=0,\,b_1=1,\,J_A=0.3$. $b_3=-1$ and  $\lambda_R=0.3$ is used for all plots.
    }
    \label{fig:dwave}
\end{figure}
(III) \underline{\emph{Altermagnet/ $s+d$-wave SC}}. Adding a small $d$-wave pairing component to limit (I) yields a fully gapped chiral TSC with $\mathcal{N}\neq 0$ without  the need for crystal anisotropy. Indeed, Fig.~\ref{fig:dwave}(a) shows that the edge spectrum for  $\Delta_0=2\Delta_1$ and $b_1=1,b_3=-1$ hosts CMFs such that $|\mathcal{N}|=1$.
While this work has focused on fully gapped TSCs, increasing the strength of $\Delta_1>\Delta_0$ creates bulk nodes in the superconducting state; the CMFs can survive as shown in Fig.~\ref{fig:dwave}(b). A fully gapped chiral TSC can also be realized in a dominantly $d$-wave pairing system coupled to an altermagnet in the presence of a distortion $b_1\neq 1$ and complex gap, $s+id$.


\blue{\emph{Majorana corner modes: SC/2D TI/Altermagnet heterostrcutures}}.---
As a final application, we demonstrate routes to higher order topology using altermagnets. As an example, we will show Majorana corner modes
in a heterostructure where an altermagnet is adjacent to a SC-proximitized 2D TI, as shown in Fig.~\ref{fig:adpic}(c) \cite{note1}.

Consider the following Hamiltonian for a normal-state 2D time-reversal invariant TI (2DTI) coupled to an altermagnet:
\begin{align}\label{2dTI}
    h_{TI}(\mathbf{k})=& A(\vex{k})+M(\vex{k})+J_A(\vex{k})
\end{align}
where $A(\vex{k})=A(\sin(k_x)\kappa^z\sigma^x+\sin(k_y)\kappa^z\sigma^y),\,M(\vex{k})=\left[B+t\left(4-2(\cos(k_x)+\cos(k_y))\right)\right]\kappa^x$, $J_A(\vex{k})=J_A(\cos(k_x)-\cos(k_y))\kappa^z\sigma^z$ and $\kappa^i$ and $\sigma^i$ are Pauli matrices that denote orbital and spin spaces. In the limit $J_A=0$ and  $0<-B<8t$, Eq.~\eqref{2dTI} describes a 2DTI exhibiting helical edges states.
Upon proximitizing with an altermagnet, i.e., turning on $J_A \neq 0$,
Eq.~\eqref{2dTI} becomes a HOTI protected by $C_4T$ symmetry that hosts single, complex fermion corner modes and fractional $e/2$ corner charges.
Such a Hamiltonian is experimentally feasible: the 2D TI layer could be, e.g., HgTe, WTe$_{2}$, bismuthene on SiC or a monolayer iron chalcogenide \cite{bernevig2006quantum,qian2014quantum,reis2017bismuthene}, and the altermagnet could be one of the materials described in Ref.~\cite{altermagnet2} such as RuO$_2$. 
\begin{figure}[tb!]
    \centering \includegraphics[width=0.48\textwidth]{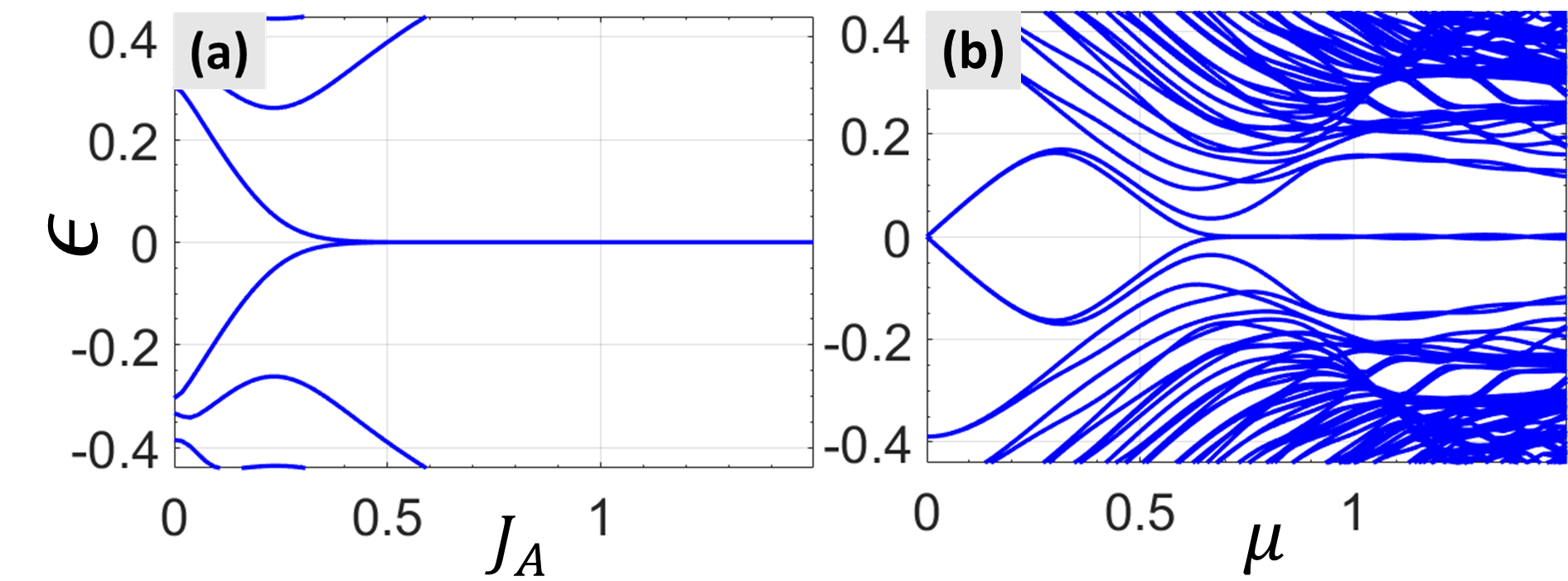}
    \caption{The corner spectrum of a 2D TI proximitzed by altermagnetism and a SC with pairing (a) $\Delta^1$ versus $J_A$ for $\mu=0$ or pairing (b) $\Delta^4$ versus $\mu$ for $J_A=0.5$. $A=1,\,B=-0.8,\,t=0.5, \Delta=0.3$ is used for both plots.}
    \label{fig:HOTISC}
\end{figure}

We now consider adding superconductivity. There are six possible antisymmetric, $k$-independent superconducting pairing terms, $\mathbf{\Delta}^i,\,i=0,\dots, 6$.
We leave a discussion of each of the pairing terms to future work, and here focus only on two options. First, upon proximitizing the HOTI phase with the simplest intra-orbital spin singlet pairing of the form $\mathbf{\Delta}^1=\kappa^0\sigma^2$, the complex fermion corner states become a pair of MZMs and gap out. However, as we found in 1D, when the chemical potential $\mu$ is tuned to zero, there is a regime where the pair of MZMs corner modes survive. Since the superconductivity competes with the altermagnetism, we expect that as the pairing increases the edges will undergo a transition between a phase gapped by altermagnetism and a phase gapped by superconductivity, the latter of which has no corner modes. To illustrate, at a fixed pairing strength, Fig.~\ref{fig:HOTISC}(a) shows that a critical strength of the altermagnetism is required to generate the pair of Majorana corner modes when $\mu=0$. Interestingly, the transition from trivial to higher order superconducting phases occurs at the edges of the sample, providing a superconducting analogue of a boundary obstructed topological phase \cite{PhysRevResearch.3.013239}.

Second, consider a spin-triplet pairing $\Delta^4=\kappa^2\sigma^1.$  Momentarily ignoring the altermagnetism, this model describes a time-reversal-invariant first-order TSC in class DIII when the chemical potential enters the bulk bands. Such a system harbors helical Majorana edge states that will immediately be gapped by altermagnetism in a spatially alternating pattern, leading to corner MZMs as shown in Fig. \ref{fig:HOTISC}(b).

\blue{\emph{Discussion}}.--- We now remark on the experimental outlook. First, while here we focused on altermagnets with $d$-wave symmetry, generalization to other symmetry groups is straightforward. Second, the wealth of both metallic and insulating altermagnet candidates \cite{altermagnet1,altermagnet2} affirms the promising nature of our proposed setup to realize TSCs with vanishing net magnetization.
Finally, we expect that the 1D and 2D platforms we proposed have interesting extensions to 3D topological phases
\cite{ghorashi2022higher, PhysRevLett.125.037001, PhysRevLett.125.266804, PhysRevB.100.020509}.

\emph{Acknowledgment}.--This work was supported by the Air Force Office of Scientific Research under Grant No. FA9550-20-1-0260. J.C. is partially supported by the Alfred P. Sloan Foundation through a Sloan Research Fellowship. The Flatiron Institute is a division of the Simons Foundation. T.L.H was supported by the Center for Quantum
Sensing and Quantum Materials, an Energy Frontier
Research Center funded by the U. S. Department of
Energy, Office of Science, Basic Energy Sciences under
Award DE-SC0021238. This research is completed during the program ``A Quantum Universe in a Crystal: Symmetry and Topology across the Correlation Spectrum'' at the Kavli Institute for Theoretical Physics, supported in part by the National Science Foundation under Grant No. NSF PHY-1748958.


\begin{thebibliography}{39}
\bibitem [{not()}]{note1}%
  \BibitemOpen
  \href@noop {} {\bibinfo  {journal} {The set-up is similar to that proposed in
  Ref.~\cite{PhysRevLett.124.227001}, but with the uniform magnetic field
  substituted by altermagnetic order.}\ }\BibitemShut {NoStop}%
\end{thebibliography}
%

\end{document}